\documentclass{PoS}
\usepackage[shortlabels]{enumitem}
\usepackage{wasysym}
\title{We are all the Cosmic-Ray Extremely Distributed Observatory}

\ShortTitle{CREDO}
\author{
	\small{
	N. Dhital$^1$, P. Homola$^1$, J. F. Jarvis$^{1,\ 2}$, P. Pozna\'{n}ski$^{1, \ 3}$, \speaker{K. Almeida Cheminant}$^1$,
	\L{}. Bratek$^1$, T. Bretz$^4$, D. Gora$^1$, P. Jagoda$^{1,\ 5}$, J. Ja\l{}ocha$^3$, K. Kopa\'{n}ski$^1$, D. Lema\'{n}ski$^{1,\ 3}$, M. Magry\'{s}$^6$, V. Nazari$^{1,\ 7}$,
	J. Niedzwiedzki$^5$, M. Nocu\'{n}$^5$, W. Noga$^1$, A. Ozieblo$^6$, 
	K. Smelcerz$^{1,\ 3}$, K. Smolek$^8$, J. Stasielak$^1$, S. Stuglik$^{1,\ 3}$, M. Su\l{}ek$^{1,\ 3}$, O. Sushchov$^1$ and J. Zamora-Saa$^7$ \\
	\llap{$^1$}Institute of Nuclear Physics Polish Academy of Sciences, Radzikowskiego 152, 31 -- 342 Cracow, Poland\\
	\llap{$^2$}School of Physical Sciences, Open University, Buckinghamshire, MK7 6AA, United Kingdom\\
	\llap{$^3$}Cracow University of Technology, Warszawska 24, 31 -- 155 Cracow, Poland\\
	\llap{$^4$}RWTH Aachen University, Physics Institute III A, Aachen, Germany\\
	\llap{$^5$}AGH University of Science and Technology, al. Mickiewicza 30, 30 -- 059 Cracow, Poland\\
	\llap{$^6$}AGH University of Science and Technology, ACC Cyfronet AGH, ul. Nawojki 11, 30-950 Cracow, Poland\\
	\llap{$^7$}Joint Institute for Nuclear Research, Dubna 141980, Russia\\
	\llap{$^8$}Institute of Experimental and Applied Physics, Czech Technical University in Prague, Horsk\'{a} 3a/22, 128 00 Praha 2, Czech Republic\\
	E-mail: \email{niraj.dhital@ifj.edu.pl}
	}
}
\abstract{The Cosmic-Ray Extremely Distributed Observatory (CREDO) is an infrastructure for global analysis of extremely extended cosmic-ray phenomena, so-called super-preshowers, beyond the capabilities of existing, discrete, detectors and observatories. To date cosmic-ray research has been focused on detecting single air showers, while the search for ensembles of cosmic-ray events induced by super-preshowers is a scientific terra incognita - CREDO explores this uncharted realm. Positive detection of super-preshowers would have an impact on ultra-high energy astrophysics, cosmology and the physics of fundamental particle interactions as they can theoretically be formed within both classical (photon-photon interactions) and exotic (Super Heavy Dark Matter particle decay and interaction) scenarios. Some super-preshowers are predicted to have a significant spatial extent  - a unique signature only detectable with the existing cosmic-ray infrastructure taken as a global network. An obvious, although yet unprobed, super-preshower `detection limit' would be located somewhere between an air shower, induced by a super-preshower composed of tightly collimated particles, and a super-preshower composed of particles spread so widely that only few of them can reach the Earth. CREDO will probe this detection limit, leading to either an observation of an as yet unseen physical phenomenon, or the setting upper limits to the existence of large extraterrestrial cascades which would constrain fundamental physics models. While CREDO's focus is on testing physics at energies close to the Grand Unified Theories range, the broader phenomena are expected to be composed of particles with energies ranging from GeV to ZeV. This motivates our advertising of this concept across the astroparticle physics community.
}
\FullConference{35th International Cosmic Ray Conference --- ICRC2017\\
		10--20 July, 2017\\
		Bexco, Busan, Korea}

\begin{document}
\vspace{-0.1cm}
\section{Introduction}
\vspace{-0.3cm}
Most of the scientific efforts undertaken so far in Dark Matter (DM) searches have revolved around the leading
paradigm in the field-- that they are weakly interacting massive particles with masses of the order of
100 GeV. Experimental efforts for the search of such DM canditates have involved both direct and indirect detection methods \cite{7, 8, 9}, as well as 
the detection of production of such particles at collider experiments \cite{10}. However, these recent efforts have
failed to produce a conclusive evidence of existence of DM candidates within the considered paradigm.
It is therefore intuitive to consider other natural scales, viz., Planck's scale and the Grand Unified Theory (GUT) scale \cite{1,2} for DM search.
Candidate dark matter particles in these regimes are believed to be super heavy ($M \geq 10^{23} \mathrm{eV}$),
and are referred to as Super Heavy Dark Matter (SHDM).
Such particles could be produced in the early Universe during the inflation phase which would decay or
annihilate presently leading to observable products, mainly photons and neutrinos in the Ultra High Energy (UHE) range \cite{11}. 
Meanwhile, even after several decades since the first observation of UHE Cosmic Rays (UHECR), many unresolved questions on their origins/sources,
propagation and composition still exist. Both of these problems constitute a complex
conundrum to physicists, answers to which will have a very important scientific impact.
A plausible solution to both problems
is that a significant fraction of UHECRs are produced as a consequence of decay or coannihilation
of SHDM particles. Proposed single solution for these two problems is in agreement with the Ockham rule which favors the least possible set of 
solutions to a collection of seemingly independent problems. However, an apparent inconsistency exists with the aforementioned assumption as the highest energy
cosmic ray events observed so far by the leading collaborations -- the Pierre Auger and the Telescope Array collaborations, are not considered photon candidates if
the present state of the art air shower reconstruction procedures are applied. In fact, there are no photon candidates within the whole energy range where
SHDM model should give photon flux, i.e. for energies above around $10^{18}$ eV
and this non-observation result leads to the very stringent upper limits to
photon fluxes \cite{erc3, erc4}. However, these analyses do not take into account the possibilities of underlying
physics which would significantly change the signature of UHE photons observed at the Earth, or the energy
spectrum of electromagnetic component arriving at the Earth. The former is attributed to the uncertainty of
interaction models in UHE regime, while the latter is attributed to possible interactions the UHE photons undergo
as they propagate in space.

\begin{figure}
\centering
\includegraphics[width=0.6\textwidth]{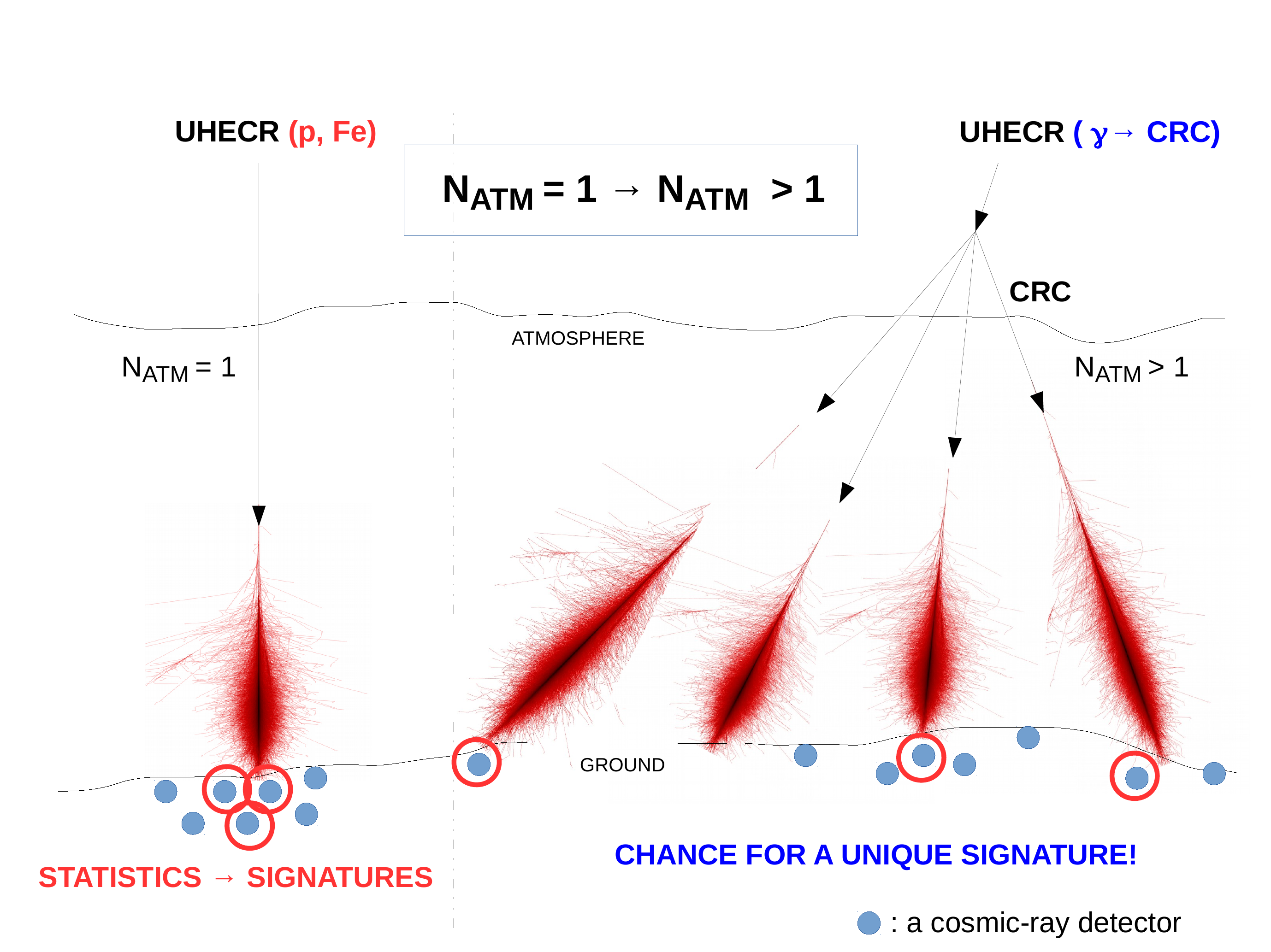}
\caption{\small A generalized schematic illustrating the novelty in the cosmic-ray research based on the consideration of cosmic rays correlated in time (right) with respect to the state-of-the art approach focused on uncorrelated particles (left).\label{fig:N1_N4}}
\end{figure}

Several theoretical models present explanation of the non-observation of UHE photons possibly produced as a consequence of aforementioned scenarios.
For example, in Lorentz Invariance Violation scenarios based on the concept of photon decay (see Ref. \cite{A} for a review on various concepts) the lifetime of a UHE photon would be extremely short ($ \ll 1$ sec), which would dramatically reduce its chances of reaching the Earth. On the other hand, one might approach observing the result of a UHE photon decay, an extensive cascade composed mainly of photons below the decay threshold. Here we consider such cascades in a general way, regardless of the primary process, and call them cosmic-ray cascades (CRC) or super-preshowers (SPS) - per analogiam to the preshower effect, a QED interaction of a UHE photon with the geomagnetic field (see e.g. Ref. \cite{17} and references therin). The novelty in this approach is illustrated in Fig. \ref{fig:N1_N4}. Within the state-of-the-art cosmic-ray research one is focused on single, uncorrelated in time cosmic rays arriving at the top of the atmosphere ( $N_\mathrm{ATM} = 1$ ), while the novel approach is oriented on a detection of cosmic-ray cascades: ensembles of cosmic rays correlated in time, ($N_\mathrm{ATM} > 1$). The latter concept includes a potential of signatures allowing an event-by-event identification of the primary particle.
Detection of SPSs have a potential
for probing unprecedented aspects of particle astrophysics. Despite this immense potential, until recently a dedicated infrastructure or network to study the SPSs 
was not in place.
With detection of SPSs, understanding physics at the UHE regime and possibly draw some meaningful inference on the DM as some of its primary goals,
the Cosmic-Ray Extremely Distributed Observatory (CREDO) was incepted in August 2016 \cite{CREDO_INAUG}.
\vspace{-0.3cm}
\section{Science Goals}
\vspace{-0.3cm}
As previously mentioned, SHDM scenarios might be directly connected to the other unsolved astrophysical
problem -- existence of UHECRs. The puzzle is not only the pure existence of particles with energies exceeding
$10^{20}$ eV. Added to the puzzle is the fact that the reconstructed directions of UHECR do not point back to
astrophysical sources \cite{12}. Also, the conclusions on the UHECR composition and spectrum cutoff from the
major experiments are not in a perfect harmony \cite{13,14} and there are general difficulties in explaining the
multi-channel data \cite{15}. The complexity of the UHECR puzzle indirectly supports an alternative scenario that
could be capable of generating UHE particles without an absolute need for correlation with the sources. While
gravitational properties of SHDM particles are rather unquestioned, the distribution of SHDM particles seems
to be disputable, e.g. not every galaxy can be an SHDM source. It is for instance not clear in the case of our
Galaxy \cite{16}. If SHDM distribution is not too far from uniform on a scale of super galaxy cluster, or if the
propagation horizon of UHE photons is more limited than we extrapolate, we might not see SHDM sources. 
On the other hand, conclusions about the constraints on the
sources or processes leading to the production of UHE photons require a number of fundamental assumptions
concerning the electrodynamics at the GUT energies \cite{20}, hadronic properties of UHE photons \cite{21}, or space-
time structure \cite{22}. Also, modeling the propagation of UHE photons through the intergalactic, interstellar and
even interplanetary medium requires assumptions difficult to verify.
In addition, we are
unsure about the physics processes at GUT energies. Moreover, we expect deviations from Standard Models
in cosmology and particle physics because of the apparent difficulties in reconcilement between theories of
General Relativity and Quantum Field Theory (see e.g. Ref. \cite{B}). Therefore, any conclusions where the GUT uncertainties are
involved, in particular the conclusions based on the upper limits to UHE photons, are disputable.
\begin{figure}
\centering
\includegraphics[width=0.7\textwidth]{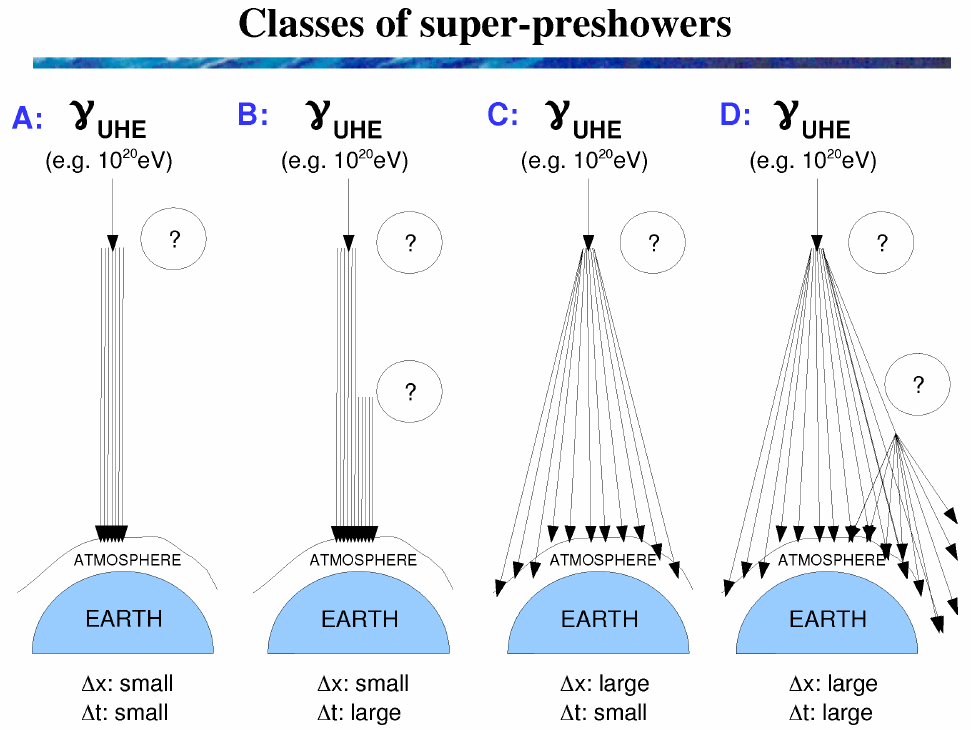}
\caption{\small A schematic of classes of ``super-preshowers''. Classification is based on widths of
spatial and temporal distribution of product particles. The question marks represent the
unknown or uncertain physical processes at UHE \cite{23}.
\label{fig:SPS_Classes}
}
\end{figure}
Physics mechanisms which lead to a cascading of most of the UHE photons before they reach the Earth
and consequently shrink their astrophysical horizon are illustrated in Fig. \ref{fig:SPS_Classes}. Given the occurence of such
mechanisms or processes in reality, UHE photons have little chance to reach the Earth, and what can be observed
on the Earth is the resulting particles most likely in a form of large electromagnetic cascades. One example of
such a cascading process is the preshower effect \cite{17} which is due to an interaction of a UHE photon and
secondary electrons with the geomagnetic field. Although this does not produce an extended cascade, a generalized notion to the
same phenomena to a UHE photon traveling towards the Earth through  close distance of the Sun can produce an extended cascade at the Earth.
Following the classification scheme in Fig. \ref{fig:SPS_Classes}, particles (mainly
photons and a few electrons/positrons) in a cascade initiated due to preshower effect in the geomagnetic field
have very narrow spatial and temporal extents and fall under class A.
\begin{figure}
\centering
\includegraphics[width=0.49\textwidth]{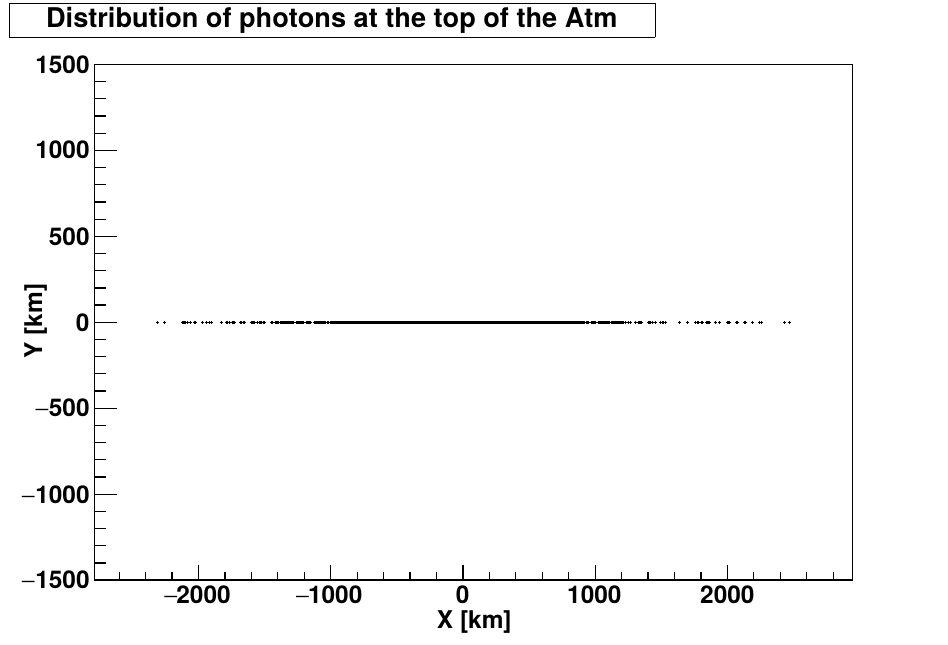}
\includegraphics[width=0.49\textwidth]{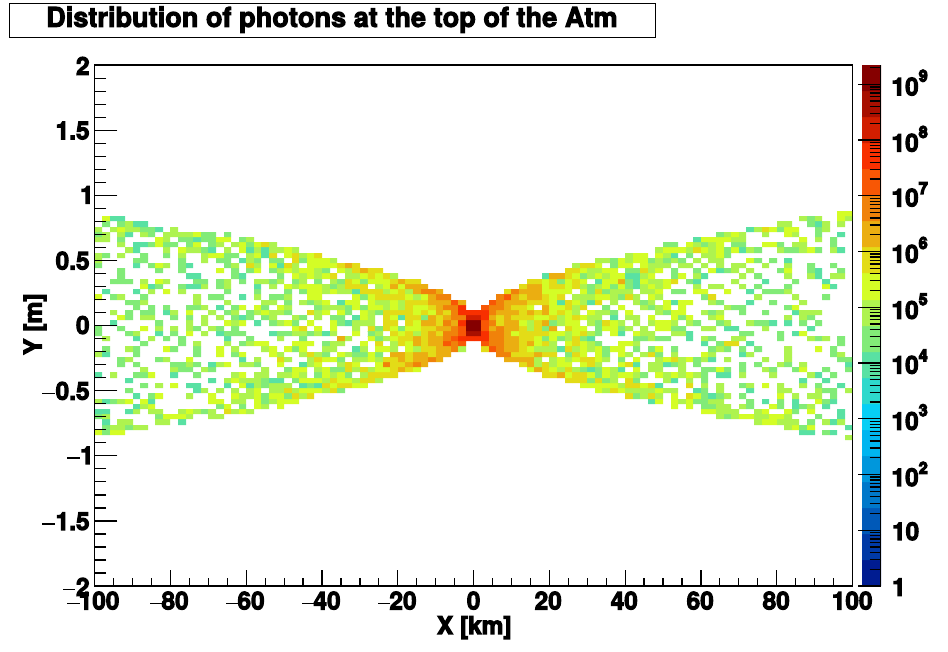}
\caption{\textit{Left panel:} Distribution of photons with $E > 10^{13}$ eV in a Sun-induced SPS at the top of the atmosphere for 10 EeV primary photon whose impact
parameter relative to the Sun is 2.5 $R_\mathrm{\astrosun}$. \textit{Right panel:} Same but zoomed-in and weighted by energy. Energies are expressed in GeV and are color-coded as
shown in the adjoining color-palette. 
\label{fig:SunSPS}
}
\end{figure}
For same effect occuring at the vicinity of
the Sun, one would expect particles with narrow time distribution but large spatial extent (see Fig. \ref{fig:SunSPS}) as the cascade arrives
at the top of the Earth's atmosphere (class ``C''). Also, other less known scenarios might cause distribution of arrival times of the particles quite extended 
with narrow spatial distribution (class ``B''), or both distributions are extended (class ``D'').

The worldwide network of cosmic ray detectors being organized under the CREDO flag
will not only be a unique tool to study fundamental physics laws,
it will also provide a number of other opportunities, including geophysics and spaceweather studies.
Among the geophysical research directions one lists the potential of CREDO to enable investigations of the correlations between the rate of low energy secondary cosmic rays and seismic effects. Existence of such correlations and their potential to forecast major earthquakes is already being discussed in the literature (see e.g. Ref.  \cite{28}). A globally distributed infrastructure like CREDO will largely increase the chances to verify the hypothesis which might lead to a way of saving human lives and properties in seismically active regions.
\vspace{-0.3cm}
\section{Detectors}
\vspace{-0.3cm}
The main goal of CREDO is the detection of very extended cascades in a global scale, in which
secondary particles have large scale time correlations.
This objective can be achieved only when CREDO is realized as a single framework of detectors spanning an area of global scale.
Since building such a detector system in its entirety is practically infeasible for several reasons such as financial and time constraints, CREDO has been incepted as a detector
framework consisting of  i) existing detectors, both educational cosmic ray detectors and the state-of-the-art cosmic ray detection facilities,
ii) new detectors which can be designed and commissioned easily and economically, and iii) common electronic devices with photo-sensors capable of cosmic ray detection using
dedicated applications via an active participation from the public. The participating detectors are interconnected via a common network for the storage of data 
into a single database, and for a real-time or a near real-time monitoring and data analysis.

As it stands now, CREDO has already reached out to a number of existing cosmic ray detection facilities for a global collaborative effort. 
Consequently, CREDO has been able to perform its first proof of concept studies with limited amount of real cosmic ray data from the following:
\begin{enumerate}[noitemsep,topsep=2pt]
\item High School Project on Astrophysical Research with Cosmics (HiSPARC) -- a network of cosmic ray detectors for educational purpose, spread across the Netherlands,
Denmark and the UK \cite{42}.
\item Shower of Knowledge -- a small network of educational detectors located on the roofs of buildings at the Joint Institute of Nuclear Research in Dubna, Russia \cite{43}.
\item QuarkNet -- A teacher professional development program which includes students perform cosmic ray research using cosmic ray detectors\cite{QN} distributed across
several countries.
\end{enumerate}

In addition to using the existing detectors, CREDO will have new cosmic ray detectors within its framework.
Installing such detectors at various locations across the globe is important as this will help us understand an overall performance of CREDO and possibly for triggering
or alerting nearby more basic but numerous detectors when needed. The number of new detectors to be deployed will depend on the financial constraints and physics requirements which
will be determined by a study to be performed in the near future.

Another component of CREDO will be common electronic devices like smartphones equipped with cameras which are capable of detecting cosmic rays. The idea of detecting cosmic rays with
smartphones has already been explored by two collaborations: the Distributed Electronic Cosmic-ray Observatory (DECO) \cite{30} and Cosmic RAYs Found In Smartphones (CRAYFIS) \cite{31}.
In addition to these two projects, a new smartphone application for cosmic ray detection is being developed within CREDO. This one will be an open source project ready to absorb the enthusiasm of users and developers across the world.
A mobile application to detect cosmic rays is an elementary, although not critical component of CREDO. Participation by a large number of science enthusiasts will naturally expand the geographic extent of CREDO and thus help reach its main goals faster.
Given a wide range of objectives including education, fundamental science goals and a possibility to perform interdisciplinary studies in
areas like spaceweather and earthquake prediction, participation from a large number of science enthusiasts is expected. Once the application is fully developed and is in use,
all relevant information regarding cosmic ray hits, locations of the detectors and the timestamps of the hits are sent to a central database server for storage and analysis.
\vspace{-0.3cm}
\section{Monitoring and Analysis}
\vspace{-0.3cm}
Data collected by each detector within the CREDO framework is stored in a central database server. In this regard, data from new cosmic ray detectors and that from
common electronic devices will be sent to the server in real-time or near real-time. Although CREDO has similar plans for data collected by other independently existing detector systems
for the near future, there will be a periodic migration of relevant data from these detector systems for now.
Once data is available in the server, the next step is to process it. Data from individual detectors are first converted into a common format, and then sorted in time and merged
into the final form for further analysis and monitoring.
\begin{figure}
\centering
\includegraphics[width=0.48\textwidth]{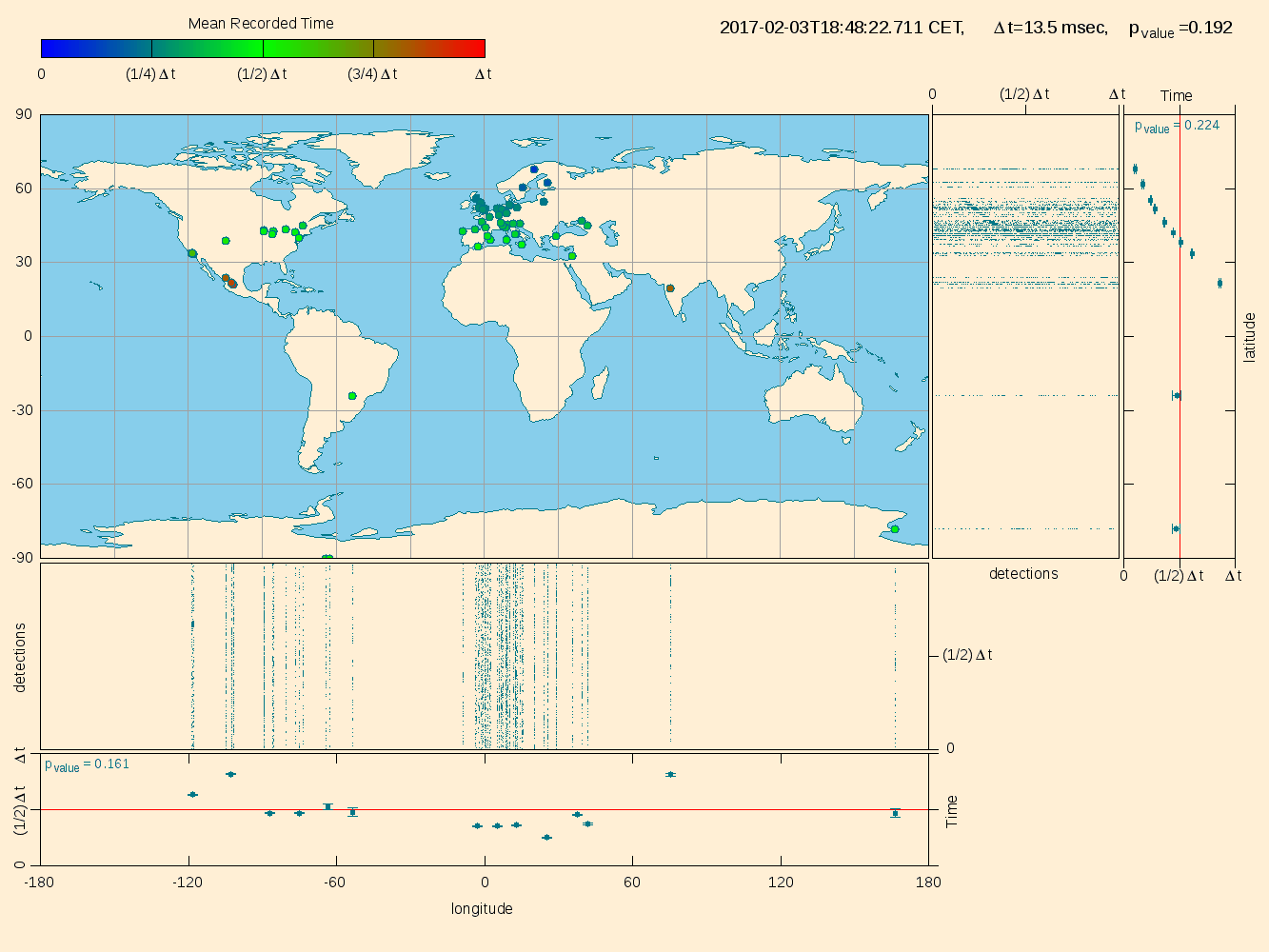}
\includegraphics[width=0.48\textwidth]{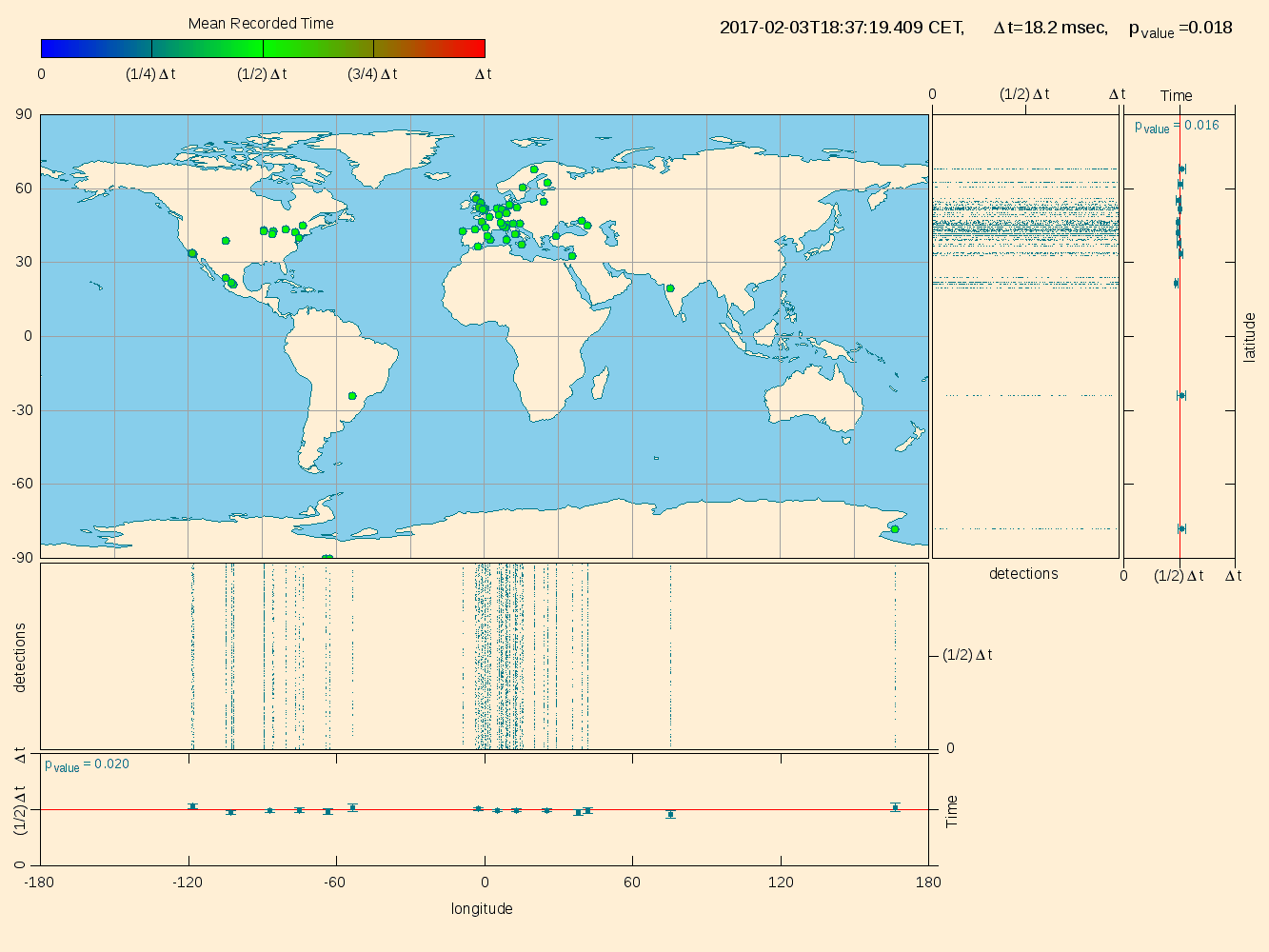}
\caption{\small An example showing average arrival time of particles for detectors distributed globally.
\textit{Left panel:} An example from SPS signal simulation. \textit{Right panel:} An example from background simulation.
\label{fig:DUW}
}
\vspace{-0.3cm}
\end{figure}
As an example, to search for a class ``C'' SPS, we look for a pattern composed of geographic locations of triggered detectors as a function of arrival times of particles
in them (Fig. \ref{fig:DUW}).
A scan performed in a small time window will give a flat line in the middle of the window for particle arrival times as a function of geographic coordinates, if recorded signals are from
purely random particles. However, if recorded signals are from the considered class of SPS, we expect a departure from this flat line. 
Based on this concept, a citizen science interface to CREDO named ``Dark Universe Welcome''  \cite{DUW}
 has been developed through which interested public, including scientists, students and science enthusiasts participate in pattern recognition and
classification of SPS events.

A number of detectors have already been integrated into the CREDO framework. Currently, data from three educational cosmic ray detectors, the HiSPARC, the Shower of Knowledge and 
QuarkNet are periodically migrated to the data storage and computing center maintained at ACC Cyfronet AGH-UST \cite{cyfronet}. An automated process is run everyday for data migration, monitoring and
analysis.
\vspace{-0.3 cm}
\section{Final remarks}
\vspace{-0.3 cm}
Despite having a strong potential for exploring unprecedented aspects of particle astrophysics with a globally distributed observatory, no such facility was in existence until recently. Our effort of an integration of existing infrastructures and facilities dedicated to cosmic ray detection has already taken a formal shape and is in an ongoing progress. Once CREDO takes its final shape, it will open a new channel to explore our Universe, namely, the SPS channel. If observed, SPSs can shed light on the interactions at energies close to the GUT scale which in turn means an unprecedented opportunity to experimentally test dark matter models and scenarios, and probe interaction models and space-time properties in otherwise inaccessible energy regime. In case of non-observation of SPS, it will valuably constrain the available theories and narrow down the area of ongoing searches for new physics. Apart from addressing fundamental physics questions CREDO will have a number of additional applications: alerting the astroparticle community on SPS candidates to enable a multi-channel data scan, potential application in interdisciplinary areas as geophysics and spaceweather, integrating the scientific community (variety of science goals, and detection techniques, wide cosmic-ray energy ranges, etc.), helping non-scientists to explore the Nature on a fundamental but comprehensible level. Since SPS photons might span a very wide range of energies, from GeV or lower to even ZeV, the subject might attract an attention of  colleagues working in seemingly separated energy regions of the particle astrophysics field. Everybody is welcome to join CREDO.
\vspace{-0.2 cm}
\section*{Acknowledgements}
\vspace{-0.2 cm}
This research has been supported in part by PLGrid Infrastructure. We warmly thank the staff at Cyfronet, for their always helpful supercomputing support.

\end{document}